\documentclass[conference, 10pt, top=0.7in]{IEEEtran}
\usepackage{amsmath,amssymb,amsfonts}
\usepackage{array}
\usepackage{graphicx}
\usepackage{textcomp}
\usepackage{xcolor}
\usepackage[utf8]{inputenc}
\usepackage{authblk}
\usepackage{graphicx}
\usepackage{bm}
\usepackage{mathtools}
\usepackage{subcaption}
\usepackage{graphicx}
\usepackage{algorithm}
\usepackage {algpseudocode}
\usepackage{acronym}
\usepackage[justification=centering, font=footnotesize]{caption}
\usepackage{xcolor}
\usepackage{cite}
\def\BibTeX{{\rm B\kern-.05em{\sc i\kern-.025em b}\kern-.08em
    T\kern-.1667em\lower.7ex\hbox{E}\kern-.125emX}}

\acrodef{RL}[RL]{Reinforcement Learning}
\acrodef{DRL}[DRL]{Deep Reinforcement Learning}
\acrodef{FL}[FL]{Federated Learning}
\acrodef{PS}[PS]{Parameter Server}
\acrodef{GNN}[GNN]{Graph Neural Network}
\acrodef{GCN}[GCN]{Graph Convolutional Network}
\acrodef{CSI}[CSI]{Channel State Information}
\acrodef{OFDM}[OFDM]{Orthogonal Frequency Division Multiplexing}
\acrodef{OFDMA}[OFDMA]{Orthogonal Frequency Division Multiple Access}
\acrodef{UL}[UL]{Uplink}
\acrodef{DL}[DL]{Downlink}
\acrodef{LoS}[LoS]{Line-of-Sight}
\acrodef{ML}[ML]{Machine Learning}
\acrodef{AI}[AI]{Artificial Intelligence}
\acrodef{NOMA}[NOMA]{Non-Orthogonal Multiple Access}
\acrodef{RRM}[RRM]{Radio Resource Management}
\acrodef{UE}[UE]{User Equipment}
\acrodef{GRL}[GRL]{Graph Reinforcement Learning}
\acrodef{IoT}[IoT]{Internet-of-Things}
\acrodef{IIoT}[IIoT]{Industrial Internet-of-Things}
\acrodef{UGRL}[UGRL]{Unsupervised Graph Representation Learning}
\acrodef{AoI}[AoI]{Age of Information}
\acrodef{SGD}[SGD]{Stochastic Gradient Descent}
\acrodef{MSE}[MSE]{Mean Square Error}
\acrodef{AP}[AP]{Access Point}

\begin{document}
\bstctlcite{IEEEexample:BSTcontrol}

\title{Uplink Scheduling in Federated Learning: an Importance-Aware Approach via Graph Representation Learning}
\author[1]{Marco Skocaj}
\author[2]{Pedro Enrique Iturria Rivera}

\author[1]{Roberto Verdone}
\author[2]{Melike Erol-Kantarci}
\affil[1]{University of Bologna, \textit{DEI} \& WiLab - National Laboratory for Wireless Communications, \textit{CNIT}, Italy}
\affil[2]{University of Ottawa, \textit{School of Electrical Engineering and Computer Science}, Canada}
\date{August 2022}

\renewcommand\Authands{ and }

\maketitle

\begin{abstract}
Federated Learning (FL) has emerged as a promising framework for distributed training of AI-based services, applications, and network procedures in 6G. One of the major challenges affecting the performance and efficiency of 6G wireless FL systems is the massive scheduling of user devices over resource-constrained channels. In this work, we argue that the uplink scheduling of FL client devices is a problem with a rich relational structure. To address this challenge, we propose a novel, energy-efficient, and importance-aware metric for client scheduling in FL applications by leveraging Unsupervised Graph Representation Learning (UGRL). Our proposed approach introduces a relational inductive bias in the scheduling process and does not require the collection of training feedback information from client devices, unlike state-of-the-art importance-aware mechanisms. We evaluate our proposed solution against baseline scheduling algorithms based on recently proposed metrics in the literature. Results show that, when considering scenarios of nodes exhibiting spatial relations, our approach can achieve an average gain of up to 10\% in model accuracy and up to 17 times in energy efficiency compared to state-of-the-art importance-aware policies.
\end{abstract}


\section{Introduction \& Motivation}
\begin{figure}[h]
\centering
\begin{subfigure}{.45\columnwidth}
\centering
  \includegraphics[width=.95\linewidth]{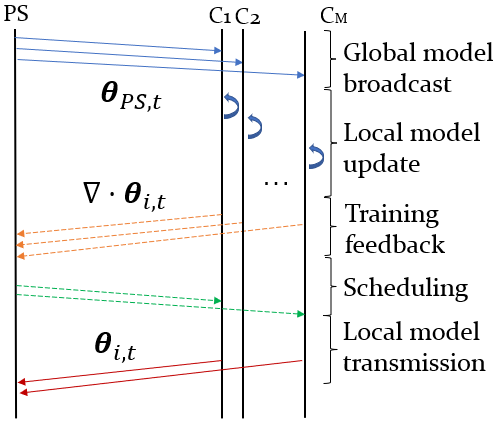}
  \label{fig:protocol1}
\end{subfigure}%
\begin{subfigure}{.45\columnwidth}
\centering
  \includegraphics[width=.95\linewidth]{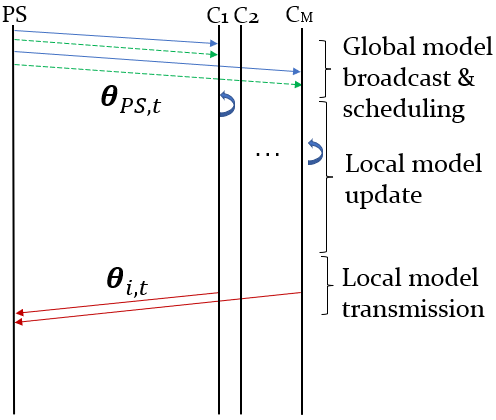}
  \label{fig:protocol2}
\end{subfigure}
\caption{Comparison between importance-aware scheduling protocols in FL. On the left side (a), classic energy-inefficient approaches require training feedback from all network clients every communication round. As the number of  clients in a network grows, this becomes highly inefficient from the computational and energetic point of view. On the right side (b), importance-aware scheduling via UGRL requires only scheduled users to perform local training.}
\vspace{-.2cm}
\label{fig:protocol}

\end{figure}

\ac{FL} \cite{mcmahan2017communication} recently emerged as a new privacy-preserving paradigm for distributed training of \ac{ML} algorithms without the need for explicit data sharing between users and a centralized computational unity. This framework is particularly appealing for next-generation 6G systems, which are foreseen to support ubiquitous \ac{AI} services and \ac{AI}-native design of users' network procedures \cite{Letayef2022Edge}. Indeed, users of a 6G network will naturally benefit from decentralized training that bypasses sharing and storing data in a centralized location. This will lead to the support of a new kind of \ac{AI}-related traffic over wireless networks: frequent exchange of \ac{ML} models introduces significant communication overhead, which raises a series of interesting novel challenges. This paved the way to a recent research area on communication-efficient \ac{FL} for 6G \cite{chen2021communication}. Wireless \ac{FL} is an example of goal-oriented communication \cite{Calvanese2021Beyond}, for which traditional \ac{RRM} methods are typically inadequate, and customized protocols must be developed \cite{hellstrom2022wireless}.\\
\noindent Arguably, one of the major challenges towards scalable and efficient 6G wireless \ac{FL} systems is the massive scheduling of user devices, from now on referred to as \textit{clients}. In fact, the central aggregation unity, namely the \textit{\ac{PS}}, generally needs to link a vast number of \ac{UE}s through a resource-constrained spectrum and thus can allow only a limited number of \ac{UE}s to send their trained weights via unreliable channels for global aggregation \cite{yang2019scheduling}. To this end, the concept of \textit{data importance}, or \textit{importance-aware communications} in FL has taken hold in the recent literature \cite{Rizk2022Federated, Rizk2021Optimal, chen2020optimal, amiri2021convergence, ren2020scheduling, Aral2020Staleness, wen2019overview}. The main idea is that by prioritizing users with high data importance, the distributed \ac{ML} training is accelerated \cite{wen2019overview}. Because explicit information on clients' data is infeasible due to privacy concerns, state-of-the-art importance-aware approaches rely on feedback information from the training of local clients' models. This approach, even though proven to be effective, has the major disadvantage of being energy and computationally inefficient. In fact, feedback-based importance-aware methods lead to the training of local models on all clients, regardless of the number of scheduled transmissions in the next communication round (Fig. \ref{fig:protocol}).

In this work, we propose a novel energy-efficient, importance-aware \ac{FL} metric based on graph representation learning, which leads to effective scheduling of client devices without any need for collecting training feedback information.

\subsection{State of the Art \& Contributions}

Over the last years, different metrics for \ac{FL} client scheduling have been proposed and discussed. In \cite{yang2020age}, by leveraging the concept of \ac{AoI}, a metric termed age of update (AoU) is introduced, which takes into account the staleness of the received parameters. In \cite{amiri2021convergence, ren2020scheduling, yang2020age, wadu2020federated} channel conditions experienced by different clients are considered during the scheduling decision. In \cite{wadu2020federated}, a channel prediction algorithm based on Gaussian process regression is incorporated into the scheduling process when dealing with imperfect channel state information. Authors from \cite{ren2020scheduling}, instead, exploit both diversity in multiuser channels and diversity in the importance of the edge devices’ local learning updates. Importance, in this case, is measured by the local parameter update's gradient divergence, which must be reported to the \ac{PS} downstream of the training of all client devices. A similar training-feedback metric of importance is introduced in \cite{amiri2021convergence}, where the significance of the model updates at the devices is captured by the L2-norm of the model update.\\
\noindent Under the frequentist setting, training data constitute a fundamental part of the inductive bias of a model. In this context, in line with the idea of importance-aware communications, knowledge about data distribution among devices would suffice for driving proper \ac{FL} scheduling decisions. However, as previously stated, this approach is unfeasible in \ac{FL} systems. Nevertheless, clients exhibit relations and correlations in a network setting, especially in the context of massive \ac{IoT} and ultra-dense 6G networks. Here, we argue that the scheduling of \ac{FL} client devices is a problem with a rich relational structure and, as a consequence, there is a need to tackle this problem effectively by taking node correlations into account. Relations among clients, which relate to local data distribution too, can be learned and inferred by encompassing network geometry and relational representation learning, while at the same time preserving users' privacy. Graphs, generally, are a representation that supports arbitrary relational structures, and computations over graphs afford a strong relational inductive bias \cite{battaglia2018relational}. By considering networks of clients as graphs, we introduce this bias in the clients' scheduling process by leveraging \ac{UGRL}. As results show, this effectively makes up for the impossibility of selecting users based on their data, while at the same time aiming for an energy and computationally-efficient scheduling protocol.\\

\noindent The main contributions of this work are listed hereafter:
\begin{itemize}
    \item We consider network geometry in the form of node embeddings obtained via \ac{UGRL} as a fundamental new metric for driving efficient \ac{FL} scheduling decisions in the context of non-i.i.d. and spatially correlated data. We aim to show it is possible to make up for the absence of explicit knowledge information about clients' data by introducing a relational inductive bias into the scheduling process.

    \item We compare the performance of the proposed scheduling metric with respect to baseline metrics recently proposed in the literature.

    \item We discuss the range of applicability of our proposed solution with respect to different kinds of data distributions with application to \ac{IoT} and 6G networks.
\end{itemize}

\section{System Model}\label{system_model}
\subsection{Network Scenario and Propagation Channel}
\begin{figure}[h]
    \centering
    \includegraphics[width=0.5\columnwidth]{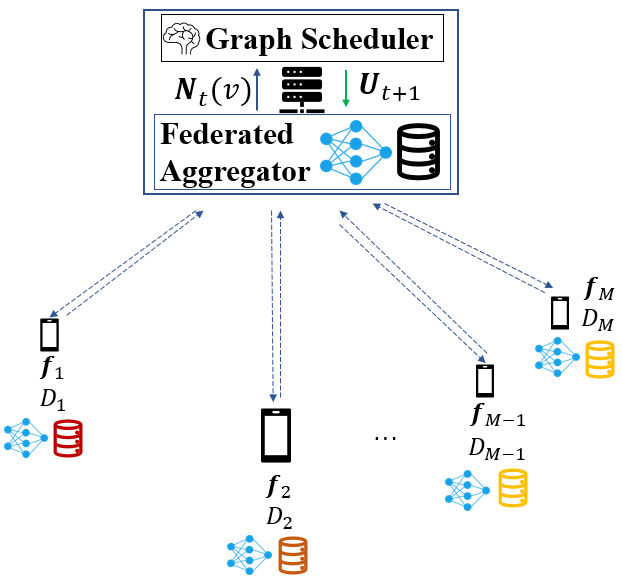}
    \caption{System Model}
    \label{fig:sys_model}
\end{figure}

Let us consider a system comprised of one \ac{AP} co-located with a \ac{PS} and multiple client devices with local data and computation capabilities, as depicted in Fig. \ref{fig:sys_model}. Physically running on the \ac{PS}, there are two software entities responsible for model aggregation and radio resource management: the \textit{Federated Aggregator} and the \textit{Graph Scheduler}, respectively described in the following subsections.

Each client $m \in \mathcal{M}$ holds a local data set $D_m = \left\{\mathbf{x}_m \in \mathcal{R}^d, \mathbf{y}_m \in \mathcal{R}\right\}$ with cardinality $|D_m|$, such that $\sum_{m \in \mathcal{M}} |D_m| = |D|$, and is equipped with a single isotropic antenna. On the \ac{PS} side, we consider an antenna with a directive  gain of 15 dBi. All clients are randomly uniformly distributed within a radius $R$ of the \ac{PS}. The \ac{PS} broadcasts the global model to the selected clients with a transmit power of 15 dBm, while the latter send their local updates with a transmit power of 10 dBm.\\
\noindent As further detailed in the next subsections, we consider the two cases of non-i.i.d. spatially correlated data and spatially correlated/uncorrelated non-i.i.d clusters of i.i.d. data. Both cases are artificially reproduced in our experiments by spatially distributing MNIST digit labels among neighbor clients.\\
\noindent Within the network area, we consider an \ac{OFDMA} scheme with perfect equalization, where  $M = |\mathcal{M}|$ clients share the same spectrum and can be assigned one of the $K < M$ set of orthogonal sub-channels for model parameters transmission. Moreover, we assume a slow fading propagation model, where each model transmission from device $m \in \mathcal{M}$ to the \ac{PS} is shorter than the channel coherence time $T_{c}$:

\vspace{-.4cm}
\begin{equation}
\small
    T_{D,m} < T_{c} \text{~~~for } m \in \mathcal{M} ,
    \vspace{-.3cm}
\end{equation}
where $T_{D,m}$ is the model transmission time for client $m$. With the aforementioned assumptions, the channel impulse response $h_{m,PS}(f,t)$ from device $m$ to the \ac{PS} loses its time and frequency dependency within a block transmission duration $T_{D,m}$ \eqref{cir}:
\begin{equation}
\small
\label{cir}
    h_{m,PS}(f,t) \rightarrow h_{m,PS} \in \mathbf{H}_{M+1} ,
    \vspace{-.1cm}
\end{equation}
where $\mathbf{H}_{M+1}$ is the channel matrix of dimension $M+1$.

\noindent Finally, we consider the Okumura-Hata model for the median path loss, and the Nakagami-m distribution for the fading propagation model, as it provides a flexible formulation to characterize Rician and Rayleigh fading. 



\subsection{Federated Learning Framework}
The goal of the \textit{Federated Aggregator} is to learn a \ac{ML} model by offloading and aggregating the training to the set $\mathcal{M}$ of distributed clients with local data. The federated training process involves a number of iterations, namely communication rounds, until convergence. Each client $m \in \mathcal{M}$, upon receiving a global model $\bm{\theta}_{PS}$ from the \ac{PS} at the beginning of a new round, executes multiple \ac{SGD} updates to minimize the model's loss function with respect to its local dataset \eqref{local_loss}:
\begin{equation}\label{local_loss}
\small
    F_m(\bm{\theta}, D_m) = \frac{1}{|D_m|} \sum_{\{x_i,y_i\} \in D_m} f(\bm{\theta}, \{x_i, y_i\}),
\end{equation}
where $F_m$ is the $m$-th client loss function and $f(\bm{\theta}, \{x_i, y_i\})$ indicates the task-dependent loss (e.g., \ac{MSE}, categorical cross-entropy, etc.) for every training example $\{x_i, y_i\}$. At every $j$-th local \ac{SGD} iteration, each client $m$ updates its local model according to \eqref{local_update}:
\begin{equation}\label{local_update}
\small
\bm{\theta}_m^{j+1}(t) = \bm{\theta}_{PS}^{j}(t) - \alpha(t)\cdot\mathbf{g}_m,
\end{equation}
where $\alpha(t)$ denotes the learning rate scheduled by the \ac{PS} for communication round $t$ and $\mathbf{g}_m := \nabla(F_m(\bm{\theta}_m, D_m)$ is the $m$-th client's gradient of the local model's weights.
Once the training is terminated, each client selected for scheduling must forward its local model $\bm{\theta}_m(t)$ to the \ac{PS}, which will update the global model upon aggregation of all received clients' models. Here, we refer to \textit{FedAvg} algorithm \cite{mcmahan2017communication}, for which the aggregation is a weighted sum described by \eqref{FedAvg}:
\begin{equation}\label{FedAvg}
\small
    \bm{\theta}_{PS}(t+1) = \frac{1}{K} \sum_{m \in \mathcal{K}}\frac{|D_m|}{\sum_{m \in \mathcal{K}}|D_m|}\cdot\bm{\theta}_{m}(t) ,
\end{equation}
where we denote by $\mathcal{K}$ ($|\mathcal{K}| = K$) the set of scheduled clients for communication.
For model evaluation, we consider a separate centralized test set $D_{test, PS}$ locally residing on the \ac{PS}.
\subsection{Graph Scheduler}
\begin{figure}[t]
    \centering    
    \includegraphics[width=.9\columnwidth]{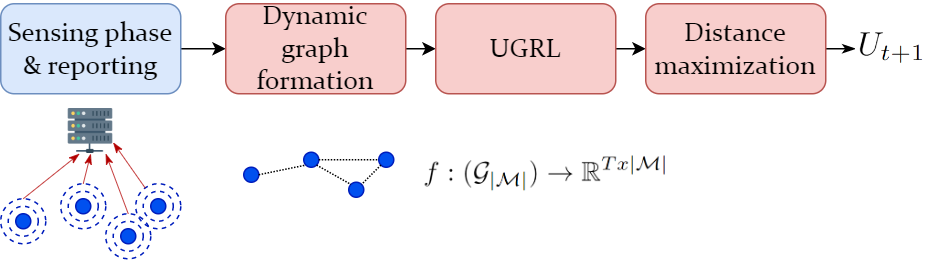}
    \caption{Graph scheduler block scheme. The scheduling sequence $\mathbf{U}_{t+1}$ is computed downstream Distance Maximization scheduler and the UGRL block. The UGRL block is responsible for the transformation $f$ from a graph space $\mathcal{G_{|\mathcal{M}|}}$ to the vectorial space $\mathbb{R}^{Tx|\mathcal{M}|}$ of node embeddings.}
    \vspace{-.5cm}
    \label{fig:graph_scheduler}
    
\end{figure}

The Graph Scheduler is the entity responsible for the procedures depicted in Fig. \ref{fig:graph_scheduler}: dynamic graph creation, UGRL, and distance maximization.

\begin{itemize}
    \item \textit{Sensing phase and reporting}: a sensing phase is performed by all client devices, and it should be repeated with a periodicity that depends on environment dynamicity. Following, all nodes $v$ report a list of sensed neighbors $N_t(v)$ (Fig. \ref{fig:sys_model}) to the graph scheduler.
    \item \textit{Dynamic graph formation}: the scheduler builds a dynamic graph from each node's list of sensed neighbors. In our approach, we retain the strongest-K neighbors, as it allows to adapt to the density of nodes' deployment. Note that alternative approaches, such as retaining adjacencies based on a received power threshold, might also be used.
    \item \textit{\ac{UGRL}}: graph representation learning involves the transformation via an encoding function $f:(\mathcal{G}_{|\mathcal{M}|}) \rightarrow \mathbb{R}^{Tx|\mathcal{M}|}$ from the graph-structured node representation $\mathcal{G}_{|\mathcal{M}|}$, to a vectorial space of T-dimensional node embeddings. In this procedure, node embeddings from $\mathcal{G}_{|\mathcal{M}|}$ are efficiently computed in an unsupervised way with the use of random walks procedures. In our scenario, we make use of Node2Vec algorithm \cite{N2V}. Further details are discussed in section \ref{algo}.
    \item\textit{Distance maximization}: the final step involves the computation of the scheduling sequence $\mathbf{U}_{t+1}$. This is based on the distance maximization between node embeddings retained in a context window of tunable dimension. Additional details are provided in section \ref{algo}.
\end{itemize}

\subsection{Data Distribution}\label{dd}

\begin{figure}[h]
\centering
\begin{subfigure}{.3\columnwidth}
  \centering
  \includegraphics[width=.95\linewidth]{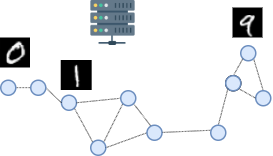}
  \caption{}
  \label{fig:sfig1}
\end{subfigure}%
\begin{subfigure}{.3\columnwidth}
  \centering
  \includegraphics[width=.95\linewidth]{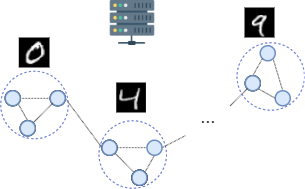}
  \caption{}
  \label{fig:sfig2}
\end{subfigure}%
\begin{subfigure}{.3\columnwidth}
  \centering
  \includegraphics[width=.95\linewidth]{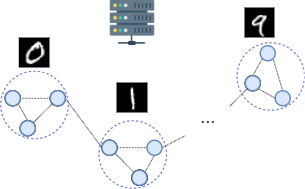}
  \caption{}
  \label{fig:sfig3}
\end{subfigure}

\caption{Typical kinds of clients data distribution in AI-native network procedures and IoT: (a) Non-i.i.d, spatially correlated data distribution, (b) Non-i.i.d, spatially uncorrelated clusters of i.i.d. data and (c) Non-i.i.d, spatially correlated clusters of i.i.d. data}
\label{fig:data_distribution}
\vspace{-0.5cm}
\end{figure}

The intuition behind the proposed approach is that the clients' data distribution reflects the structural relation of nodes in a graph. Consequently, this method does not apply to the trivial case of i.i.d. data. Vice versa, it is possible to think of a plethora of applications and AI-driven network procedures foreseen for 6G in which this condition holds: localization, tracking, integrated sensing and communication, channel estimation and measures of a physical quantity from a sensors network is just a non-exhaustive list of examples where clients' data are non-i.i.d. and spatially correlated. Another vertical of great importance for future 6G networks is \ac{IIoT}, where typically nodes are arranged into clusters, and nodes within a cluster might hold similar kinds of measurements (e.g., monitoring sensors inside automatic machines in a warehouse). To this end, we consider in this work typical kinds of client data distributions that find use in many real-world 6G applications:
\begin{itemize}
\item Non-i.i.d. and spatially correlated data distribution (Fig. \ref{fig:sfig1}).
\item Spatially correlated/uncorrelated clusters of i.i.d. data (Fig. \ref{fig:sfig2}, \ref{fig:sfig3}).
\end{itemize}


For benchmarking purposes, we reproduced the three scenarios described above in our simulations by distributing MNIST data arranged by labels to a set of randomly distributed clients. According to the considered scenario, clients are distributed, at the beginning of every new simulation, a random number of examples according to: a) their respective position to the PS (for the case depicted in Fig. \ref{fig:sfig1}), or b) their belonging cluster (for the case depicted in Fig. \ref{fig:sfig2}, \ref{fig:sfig3}).

\section{Problem Formulation}\label{formulation}

The observation space can be represented as a graph composed of nodes (client devices), edges (adjacency matrix), and node features $\mathcal{F}$ (\ac{FL} metrics). In the most general formulation, each edge can also be associated with a weight $|h_{i,j}|$, corresponding to the module of the complex channel impulse response $h_{i,j}$ between client $i$ and $j$, drawn from a $(M+1)$-dimensional channel matrix $\mathbf{H}_{M+1}$. Nevertheless, considering the case of orthogonal resources assignment (i.e., no inter-users interference), and assuming link reciprocity, allows for a simplification of the problem formulation, since matrix $\mathbf{H}_{M+1}$ reduces to an M-dimensional vector $\mathbf{H} = [|h_{1, PS}|, \ldots, |h_{M, PS}|]$. Hence, its elements can be represented as node features instead of edges.
\vspace{-.2cm}

\begin{equation}
\small
    S_t =
    \begin{cases}
    \begin{aligned}
      \mathbf{F} = \bigr[(\mathcal{F}_1, |&h_{1,PS}|),...,\\&(\mathcal{F}_M, |h_{M,PS}|)\bigr],
      \end{aligned}
      & 
      \begin{aligned}
      \text{with } |h_{i,PS}| \in \mathbb{R}
      \end{aligned}\\
      \\
      \mathbf{A} =
        \begin{bmatrix}
            a_{11} & \ldots & a_{1M}\\
            \vdots & \ddots & \vdots\\
            a_{M1} & \ldots & a_{MM}
        \end{bmatrix}_M, & \text{with } a_{ij} \in \{0,1\}
    \end{cases}
\label{eqn:state_space}
\end{equation}
\normalsize
\noindent In \eqref{eqn:state_space}, $\mathbf{F}$ is the feature space of every node, which includes the \ac{FL}-metrics $\mathcal{F}$ used for decision-making during the scheduling procedure, and $\mathbf{A}$ is the adjacency matrix, which is obtained downstream of the dynamic graph formation block of Fig. \ref{fig:graph_scheduler}.


\noindent The feature space is composed of the following metrics:
\begin{itemize}
    \item Age of Information (AoI): a scalar indicator, introduced in \cite{yang2020age}, describing the number of rounds that elapsed since the client was last scheduled for model transmission.
    \item Path loss $|h_{i,j}|$: the path loss value in dB related to the client-\ac{PS} link, assuming link reciprocity.
    \item L2-norm of model update $||\bm{\theta}_{i, t+1} - \bm{\theta}_{PS, t}||_2$: an importance metric indicating the L2 norm of the $i$-th client model update.
\end{itemize}

\noindent Notably, no explicit informative content can be collected about the data of the clients, as this would violate the privacy-preserving nature of \ac{FL}.

\section{Proposed Algorithm}\label{algo}
In the formulation above, nodes connected by edges have similar data, but not necessarily similar FL metric features. In fact, \ac{AoI}, for instance, does not show any spatial correlation property among nodes, as it purely depends on the scheduling mechanism. In turn, this, together with the inability to have explicit features about nodes' data, depicts a situation where client features don't necessarily reflect the structure of the graph. Moreover, the nature of the problem is naturally unsupervised, as nodes don't have any label. To this end, we make use of \ac{UGRL} via random walks in place of common supervised methods with \ac{GNN}s to incorporate information about the structure of the graph into the decision-making process. By employing Node2Vec \cite{N2V}, we are able to compute the graph encoding function $f(\mathcal{G})$ without the assistance of any node labels and features, but rather by maximizing the log-likelihood of the 2nd order biased random walks $N_s(u)$ conditioned by $f(u)$, for $u \in \mathcal{G}$ as per \eqref{N2V} \cite{N2V}.

\begin{equation}
\small
\label{N2V}
    \arg \max_f \sum_{u \in \mathcal{G}(V)} \log\left(\text{Prob}(N_s(u)|f(u)\right)
\end{equation}
\normalsize

\subsection{Distance Maximization via UGRL}
Once $f(\mathcal{G})$ is obtained, clients' scheduling relies on the distance maximization of the nodes in the embedding space. This has the effect of increasing data heterogeneity of the client devices during consecutive communication rounds. The distance among a pair of nodes $(v, u)$ in the graph $\mathcal{G}$ is evaluated as the normalized dot product (cosine similarity) $s(v, u)$ of their node embeddings $\mathbf{z}_v, \mathbf{z}_u$ \eqref{cos_sim}:

\begin{equation}\label{cos_sim}
\small
    s(v,u) = \frac{\mathbf{z}_v^T\cdot\mathbf{z}_u}{\Vert \mathbf{z}_v \Vert \cdot \Vert \mathbf{z}_u \Vert}.
\end{equation}
To introduce memory of the past actions in the scheduling process, nodes scheduled in previous communication rounds are stored in a context window $W_L$ of tunable length $L = \max(|W_L|)$. Accordingly, the similarity scores of the nodes $v \in W_L$ are computed and collected in a matrix $\mathbf{S}_{v,u}$ of dimension $|W_L| \times (M-|W_L|)$:

\begin{equation}\label{sim_matrix}
\small
\mathbf{S}_{v,u} = 
    \begin{bmatrix}
        s(v_1,u_{L+1}) & \ldots & s(v_1,u_M)\\
        \vdots & \ddots & \vdots\\
        s(v_L,u_{L+1}) & \ldots & s(v_L,u_M)
    \end{bmatrix}.
\end{equation}
A scheduling decision is finally determined by equation \eqref{dist_max}.

\begin{equation}
\label{dist_max}
\small
\arg\underset{u \in \mathcal{G} \setminus \{v \in W_L\}}{\min}\underset{v \in W_L}{\sum}s(v,u).
\end{equation}
The latter is equivalent to summing all elements of the matrix $\mathbf{S}_{v,u}$ by column, and selecting the next scheduled client as the argument corresponding to the column holding the minimum sum value, i.e., the node with maximum distance with respect to all previously scheduled nodes in $W_L$.




\section{Simulation Methodology}\label{methodology}
Experiments and evaluation were conducted on a simulator based on Tensorflow Federated Core API. The logical steps of the designed \ac{FL} framework, namely \textit{"FederatedEnv"}, are reported in Algorithm \ref{algo:communication_round}.


\begin{algorithm}[b]
\caption{FederatedEnv: logical steps}
\label{algo:communication_round}
    \begin{algorithmic}
    \footnotesize
    \For {each episode}
        \For {$m \in \mathcal{M}$}
            \State $
            \begin{aligned}             
            (r_m, \phi_m) = \left(r_m\sim U(0,R), \phi_m\sim U(0,2\pi) \right)
            \end{aligned}$
            \State$
            \begin{aligned}
                |D_m| = |D| \cdot (u_m \sim U(0, 1)/\sum u_m)
            \end{aligned}$
            \State $
            \begin{aligned}
                D_m = \text{sample\_data}(D, |D_m|, \phi_m)
            \end{aligned}$
        \EndFor
        \For {each round $t$ in $T$ and for every $m \in \mathcal{M}$}
        \State \textbf{Step 1: DL Channel State Estimation}
            \State $
            \qquad |h_{PS,m}|_{dB} = - \text{PL}_m + s \sim N(0, \sigma_s) + \mathfrak{f} \sim \mathfrak{F}(k,\omega)
                $
            \State $
            \qquad \gamma_{DL,m} = P_{tx,DL} + G_{tx} + |h_{PS,m}|_{dB} - N_{dB}
                $
        \State \textbf{Step 2: Model Broadcast}
            \State $
                \qquad \bm{\theta}_m(t) \leftarrow \bm{\theta}_{PS}(t) + \mathbf{n}\sim N(0,\sigma_{\propto \gamma_{DL,m}})
                $
        \State \textbf{Step 3: Client Update}
            \State $
            \begin{aligned}
                \qquad\bm{\theta}_m^{j+1}(t) = \bm{\theta}_m^j(t) - \alpha(t) \cdot \mathbf{g}_m
            \end{aligned}$
        \State \textbf{Step 4: UL Channel State Estimation}
           \State $
            \qquad |h_{m,PS}|_{dB} = - \text{PL}_m + s \sim N(0, \sigma_s) + \mathfrak{f} \sim \mathfrak{F}(k,\omega)
                $
            \State $
            \qquad \gamma_{UL,m} = P_{tx,UL} + G_{tx} + |h_{m,PS}|_{dB} - N_{dB}
                $
           \State \textbf{Step 5: Client Scheduling}
            \State $
            \qquad \bm{\Theta}'_{\text{M}} = \bm{\Theta}'_{\text{M}}\cdot \mathbf{U}^T = [\bm{\theta}'_1,\ldots,\bm{\theta}'_M]^T \cdot [u_1,\ldots,u_M]
            $
        \State \textbf{Step 6: Model update}
            \State $
                \qquad\bm{\theta}_{PS}(t+1) = \frac{1}{K} \sum_{m \in \mathcal{K}}\frac{|D_m|}{\sum_{m \in \mathcal{K}}|D_m|}\cdot\bm{\theta}_{m}(t) + \mathbf{n}_m
            $
        \State \textbf{Step 7: Performance Evaluation}
            \State $
            \qquad f(D_{test} | \bm{\theta}_{PS}(t+1))
            $
        \EndFor
    \EndFor
    \end{algorithmic}
\end{algorithm}
\noindent After the initialization of clients' positions $(r_m, \phi_m)_{M}$ and datasets $D_m$, the algorithm loops over a fixed number of communication rounds. Each round can be subdivided into 7 logical steps:

\begin{figure*}[t]
\centering
\begin{subfigure}{.55\columnwidth}
  \centering
  \includegraphics[width=\linewidth]{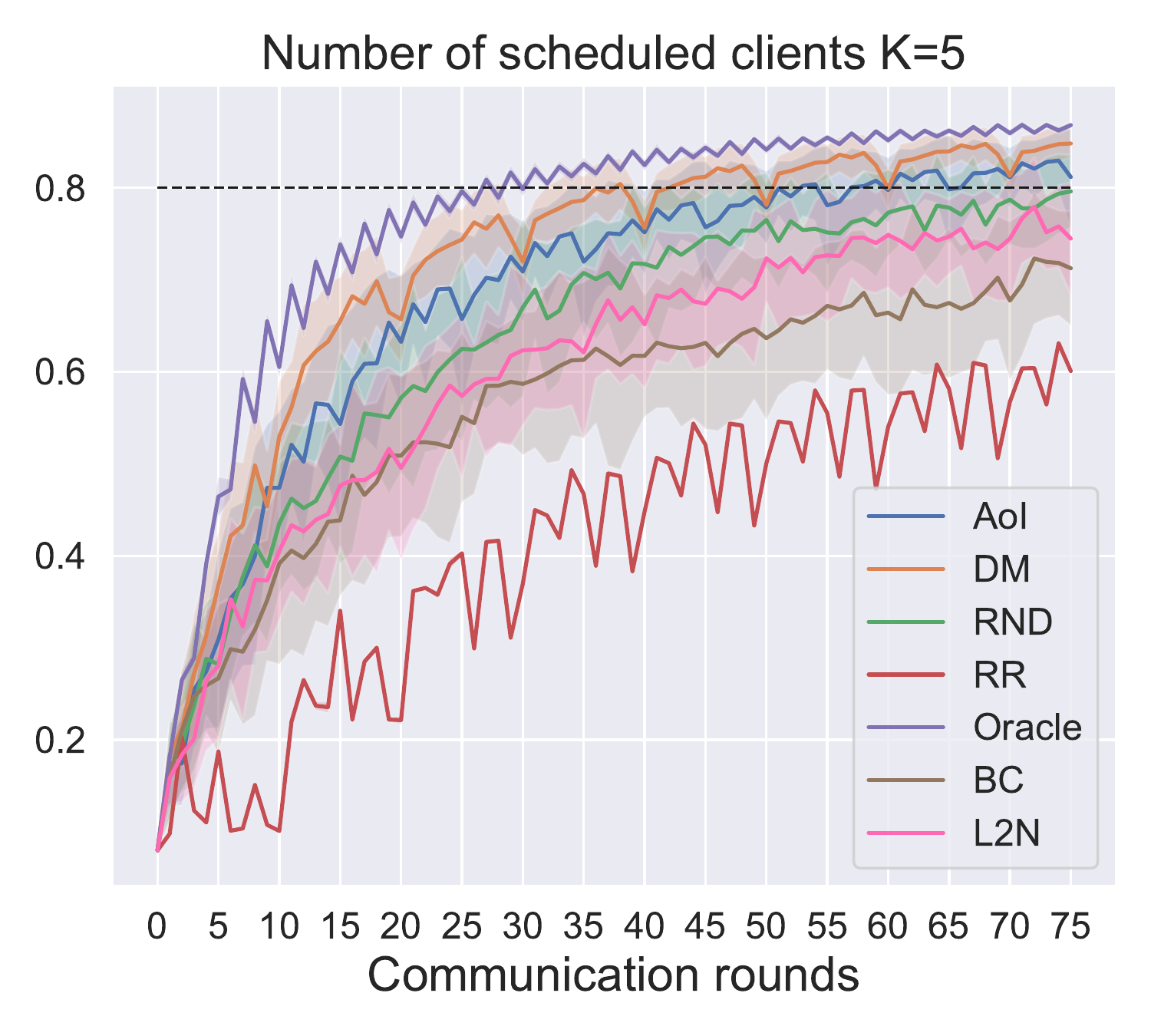}
  \label{fig:result1}
\end{subfigure}%
\begin{subfigure}{.55\columnwidth}
  \centering
  \includegraphics[width=\linewidth]{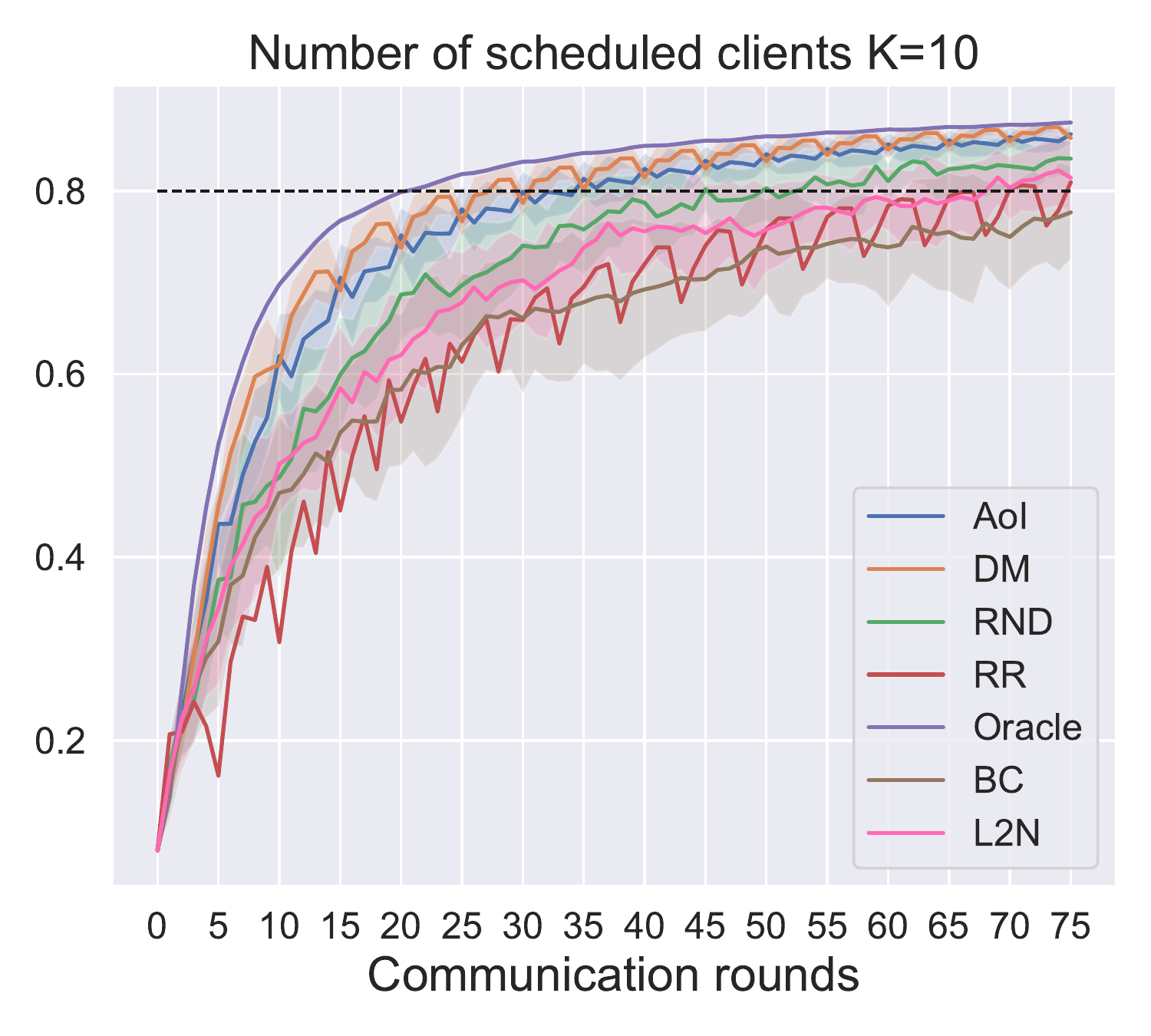}
  \label{fig:result2}
\end{subfigure}%
\begin{subfigure}{.55\columnwidth}
  \centering
  \includegraphics[width=\linewidth]{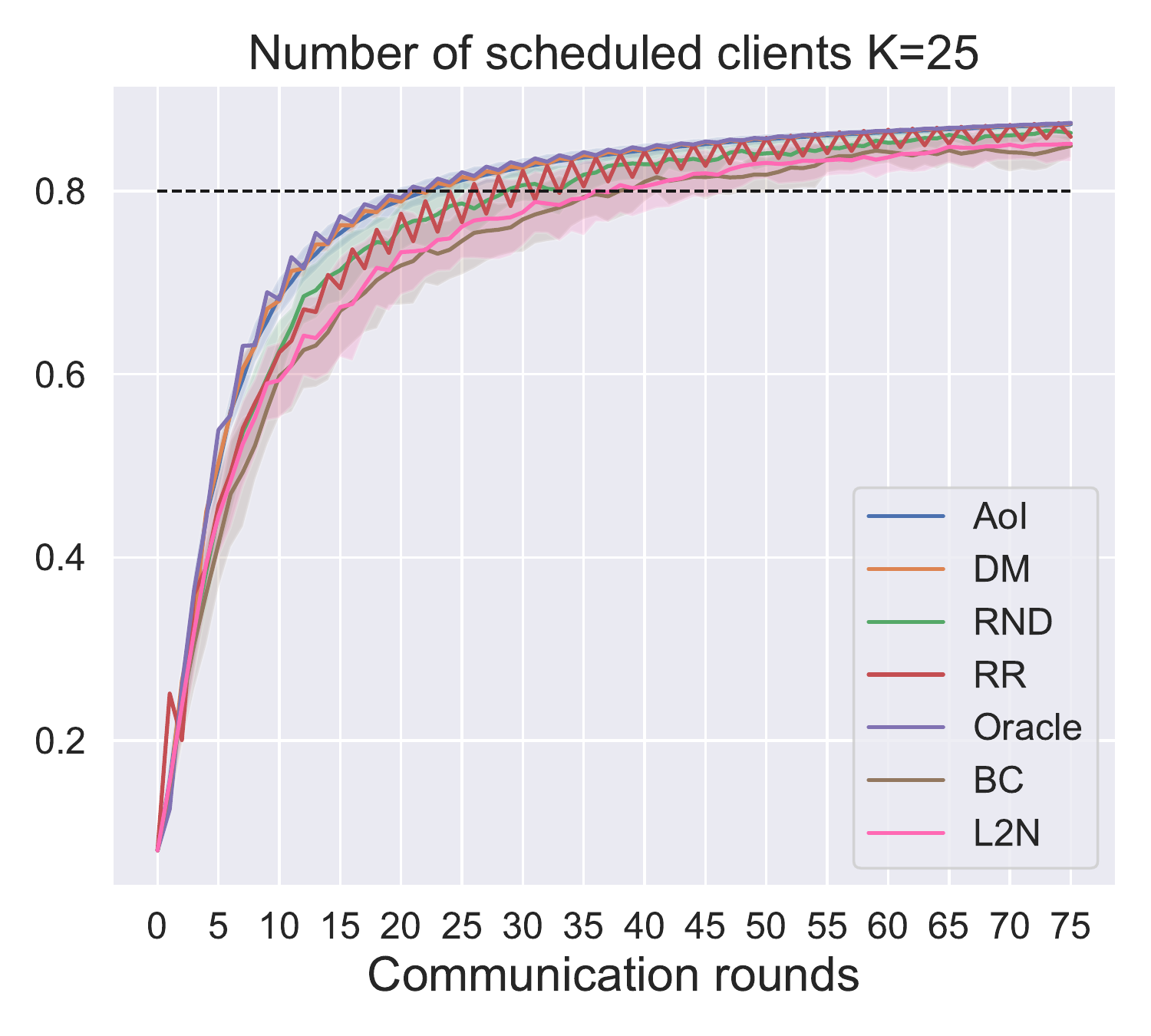}
  \label{fig:result3}
\end{subfigure}
\vspace{-0.5cm}
\caption{Training results: sparse categorical cross accuracy vs number of communication rounds for different number of schedulable users per round: (a) $K=5, M=50$, (b) $K=10, M=50$ and $K=25, M=50$}
\label{fig:results1}
\end{figure*}

\begin{enumerate}
    \item The module of the client-\ac{PS} \ac{DL} channel impulse response $|h_{PS,m}|$ is computed, assuming perfect CSI, by summing fading and shadowing contributions $(f\sim\mathfrak{F}, s\sim N)$ to the median path loss $\text{PL}$. The signal-to-noise ratio $\gamma$ is thereby computed with a noise floor $N_{dB}$ of -115 [dBm] and an \ac{AP} gain $G_{tx}$ of 15 [dBi].
    \item The \ac{PS} model $\bm{\theta}_{PS,t}$ is broadcasted to the clients. At the receiver side, Gaussian noise with standard deviation $\sigma_{DL} = \sqrt{E[\mathbf{N^2}]} = \sqrt{E[\bm{\theta}^2]/\gamma_{DL,m}}$ is added to the \ac{PS}'s model weights, where $\bm{\theta}$ is the discrete signal of weights of every model's layer. 
    \item Each client performs a local update of its model weights, controlled by the learning rate $\alpha(t)$, using stochastic gradient descent for $N_{ep}$ epochs, as per \eqref{local_update}.
    \item After local model updates, a new round of \ac{UL} channel estimation is performed in the same way as for step 1, and the corresponding \ac{UL} signal-to-noise ratio $\gamma_{UL}$ is computed.
    \item A binary mask of the scheduled users $\mathbf{U} = [u_1, \dots, u_M]$ is applied to the vector of updated model weights $\bm{\Theta}'_\text{M} = [\bm{\theta}'_1,\ldots,\bm{\theta}'_M]^T$. During the aggregation phase, only models belonging to scheduled clients will be retained and included in the aggregation process.
    \item The noisy models from the scheduled clients are aggregated as per \eqref{FedAvg}.
    \item The model is evaluated on a test set. We denote by $f(D_{test} | \bm{\theta}_{PS}(t+1))$ any metric computed over $D_{test}$.
    
\end{enumerate}

\section{Results}\label{results}

\noindent In this section, we present and comment on the obtained results. We consider a scenario where $M=50$ clients are randomly distributed among equally populated clusters (Fig. \ref{fig:sfig2}, \ref{fig:sfig3}). The distance maximization scheduling was tested against baseline policies (Table \ref{tab:baseline_policies}) based on the feature nodes metrics introduced in section \ref{formulation}. All policies have been tested on MNIST classification when transmitting a shallow neural network with 1 hidden layer, achieving 0.91 accuracy in a centralized setting.

\begin{table}[h]

\begin{tabular}{|l|l|}
  \hline
     \textbf{POLICY} & \textbf{DESCRIPTION}.\\
  \hline
     Max Age of Information (AoI)& nodes with max \ac{AoI}\\
  \hline
     Random (RND)& nodes chosen randomly\\
  \hline
     Round robin (RR)& nodes chosen in a round-trip
fashion\\
  \hline
     Best Channel (BC)& nodes with the best channel condition\\
  \hline
     Oracle (OCL) & explicit data knowledge information -\\& maximize label heterogeneity\\
  \hline
   Max L2-norm (L2N)& nodes with maximum L2-norm of\\& their local model update\\
  \hline
     Distance maximization (DM)& our proposed scheduling policy\\
  \hline
\end{tabular}
\caption{\small{Baseline policies}}
\label{tab:baseline_policies}
\vspace{-0.3cm}
\end{table}

\begin{figure}[h]
\centering
\begin{subfigure}{\columnwidth}
  \centering
  \includegraphics[width=.8\linewidth]{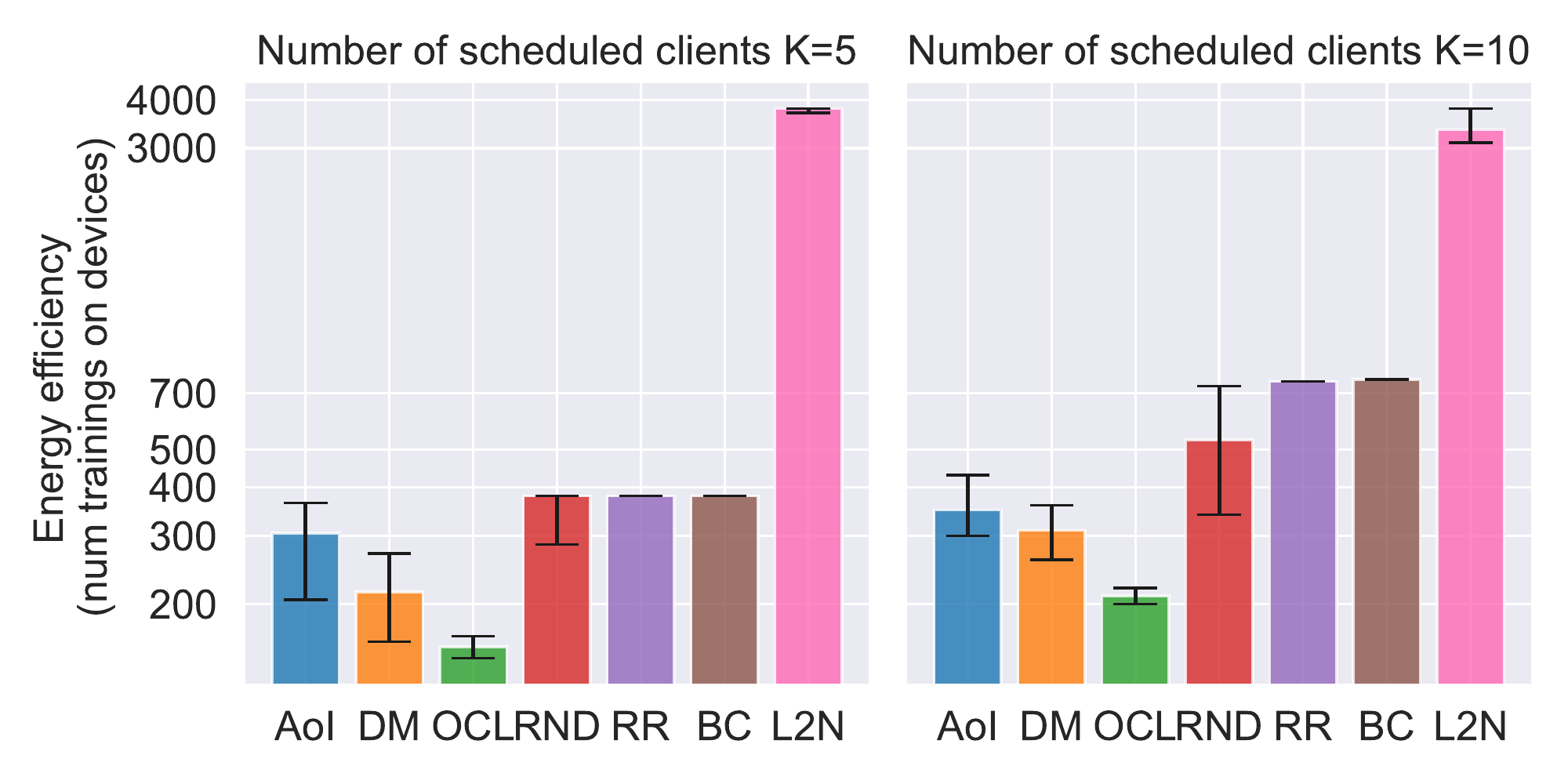}
  \label{fig:eneff1}
  \caption{Energy efficiency}
  \label{eneff}
\end{subfigure}
\begin{subfigure}{\columnwidth}
  \centering
  \includegraphics[width=.8\linewidth]{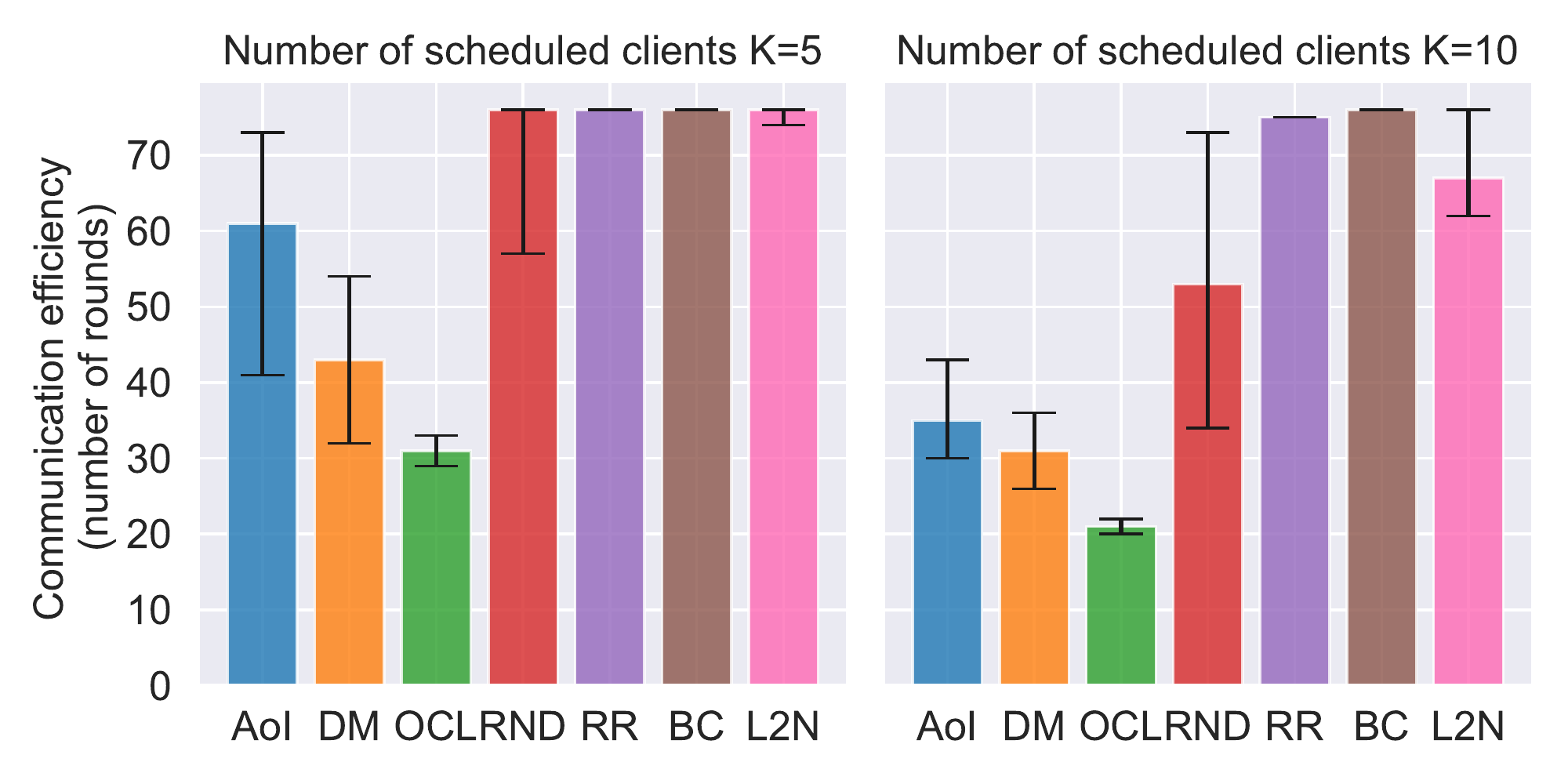}
  \label{fig:eneff2}
  \caption{Communication efficiency}
  \label{commeff}
\end{subfigure}
\caption{Training results: energy (a) vs communication efficiency (b).}
\label{fig:results2}
\end{figure}

\noindent Fig. \ref{fig:results1} and Fig. \ref{fig:results2} show performance comparison in terms of training accuracy and energy (Fig. \ref{eneff}) vs. communication efficiency (Fig. \ref{commeff}), respectively. The three metrics are evaluated as follows:
\begin{itemize}
    \item\textit{Accuracy}: the sparse categorical cross accuracy on a centralized test set.
    \item\textit{Communication efficiency}: the number of communication rounds $r$ to achieve an accuracy of 0.8.
    \item\textit{Energy efficiency}: the number of rounds $r$, multiplied by the number of training devices per round ($K$, or $M$, depending on the policy).
\end{itemize}

\noindent In all three metrics, the proposed solution outperforms the baselines and approaches the performance of the oracle, which is an empirical upper bound. In particular, the performance gap between the proposed solutions increases as the number of schedulable users ($K$) decreases. It is of particular interest to compare our proposed DM policy with respect to the L2N importance-aware policy. Since the former does not require any form of training feedback (Fig. \ref{fig:protocol}), it is much more energy efficient, as the number of clients training per round is reduced from $M$ to $K$ (i.e., only the scheduled users perform local training). Moreover, results show that, in the proposed scenario, DM outperforms L2N even in terms of model accuracy and communication efficiency (Fig. \ref{fig:results1} and \ref{commeff}). In fact, even though the L2N policy is successful in scheduling the users holding the most significant models every round (i.e., those contributing more significantly to the global model, according to the L2-norm of the local updates \cite{amiri2021convergence}), when nodes show spatial relation, this may result in the selection of nearby clients in space holding similar data. On the opposite, our proposed policy aims to achieve data heterogeneity by maximizing nodes' distance in the graph space, making it more suitable for the considered scenario. For the same reason, policies such as \ac{AoI} and random scheduling achieve better performance than round-robin, since they inherently increase data heterogeneity among the clients scheduled every round. Results show that DM can achieve an average 10\% gain in model accuracy with respect to L2N at the end of the training while increasing energy efficiency by 17 times for $K=5$. Moreover, we register an average gain of $4\%$ accuracy at the end of the training and $31\%$ in communication and energy efficiency with respect to the second-best performing policy (\ac{AoI}). Finally, it is interesting to notice how the choice of $K$ generates a tradeoff between communication and energy efficiency. Indeed, for our considered model and simulation parameters, if the aim of the designer is to maximize the system's communication efficiency, then $K=10$ yields a gain of $27,9\%$ in communication efficiency, but a loss of $30,7\%$ in energy efficiency with respect to $K=5$. Therefore, this is an indicator that for energy-sensitive applications, like \ac{IoT}, the maximization of the number of schedulable clients is not always the best design choice.
\vspace{-.3cm}

\section{Conclusion}

\noindent In this study, we present a novel metric for the scheduling of client devices in \ac{FL} applications leveraging the use of \ac{UGRL}. With respect to state-of-the-art importance-aware scheduling methods, our solution does not require any training feedback from client devices. Hence, it provides a much more computationally and energy-efficient solution. Our results indicate that, when tested against baseline importance-aware policies, our solution achieves a gain of up to $10\%$ in model accuracy, while requiring up to 17 times fewer local training phases on client devices.
\vspace{-.2cm}

\bibliographystyle{IEEEtran}

\bibliography{IEEEabrv,StringDefinitions,references}

\end{document}